\documentclass[proof]{pasj00}

\begin{document}
\SetRunningHead{Author(s) in page-head}{Running Head}

\title{The Ejecta Distributions of the Heavy Elements in the Cygnus Loop}

\author{Hiroyuki \textsc{Uchida}\altaffilmark{1}, Hiroshi \textsc{Tsunemi}\altaffilmark{1}, Satoru \textsc{Katsuda}\altaffilmark{1,2}, Masashi \textsc{Kimura}\altaffilmark{1}, Hiroko \textsc{Kosugi}\altaffilmark{1}} %
\altaffiltext{1}{Department of Earth and Space Science, Graduate School of
  Science, Osaka University, Toyonaka, Osaka 560-0043, Japan}
\altaffiltext{2}{Code 662, NASA Goddard Space Flight Center, Greenbelt, MD 20771}
\email{uchida@ess.sci.osaka-u.ac.jp}


\maketitle

\begin{abstract}
We analyzed the metal distribution of the Cygnus Loop using 14 and 7 pointings observation data obtained by the \textit{Suzaku} and the \textit{XMM-Newton} observatories. The spectral analysis shows that all the spectra are well fitted by the two-$kT_e$ non-equilibrium ionization plasma model as shown by the earlier observations. From the best-fit parameters of the high-$kT_e$ component, we calculated the emission measures about various elements and showed the metal distribution of the ejecta component. We found that the distributions of Si and Fe are centered at the southwest of the geometric center toward the blow-out region. From the best-fit parameters, we also estimated the progenitor mass of the Cygnus Loop from our field of view and the metal rich region with a radius of 25\,arcmin from the metal center. The result from the metal circle is similar to that from our entire FOV, which suggests the mixing of the metal. From the results, we estimated the mass of the progenitor star at 12-15\MO.
\end{abstract}

\section{Introduction}\label{sec:intro}
The supernova (SN) explosions blow out the various heavy elements generated by the nucleosynthesis process inside the progenitor stars. Meanwhile, the blast wave formed by the SN explosion sweeps up and heats the interstellar matter (ISM). In this way, the abundance of the progenitor star or the ISM are measured from the observation of the supernova remnant (SNR).
 
The Cygnus Loop is one of the brightest SNRs in the X-ray sky. Its age is estimated to be $\sim$10,000 yr and the distance is comparatively close to us (540\,pc; \cite{Blair05}). The large apparent size ($2^\circ.5\times3^\circ$.5; \cite{Levenson97}) enables us to study the plasma structure of the Loop.

Although the Cygnus Loop is an evolved SNR, a hot plasma is still confined inside the Loop (\cite{Tsunemi88}; \cite{Hatsukade90}). \citet{Miyata98} observed the Loop with the Advanced Satellite for Cosmology and Astrophysics (\textit{ASCA}), and detected strong highly-ionized Si-K, S-K, and Fe-L lines near the center of the Cygnus Loop. They concluded that a hot plasma, a ``fossil'' of the supernova explosion, is left in the core of the Loop. Tsunemi et al. (2007) observed the Cygnus Loop along the diameter from the northeast (NE) to the southwest (SW) with \textit{XMM-Newton} and studied the radial plasma structure. From the spectral analysis, they showed that the Cygnus Loop consists of two components with different temperatures and metal abundances. They concluded that the low-$kT_e$ component originating from the cavity-wall component surrounds the high-$kT_e$ ejecta component. In addition, they measured the metal abundances of the high-$kT_e$ component and showed the metal distribution of the ejecta. The results indicate that the abundances are relatively high ($\sim$5 times solar) and each element is non-uniformly distributed: Si, S and Fe are concentrated in the inner region while the other elements such as O, Ne and Mg are abundant in the outer region. \citet{Katsuda08} and \citet{Kimura}\footnote{The data available at http://arxiv.org/pdf/0810.4704v1} expanded the observations southward and northward respectively with the \textit{Suzaku} observatory in 7 and 10 pointings, and examined the plasma structure in their Field of View (FOV). \citet{Katsuda08} divided their FOV into 119 cells and, performed the spectral analysis. They showed that Si and Fe are more concentrated in the south part than that in the north part. This indicates a clear asymmetric structure of the metal abundances. \citet{Kimura} divided their FOV into 45 rectangular regions from NE to SW and also showed the asymmetric distribution of each heavy element; the ejecta of O, Ne, Mg are distributed more in the NE, while those of Si, Fe are distributed more in the SW. \citet{Tsunemi07} and \citet{Kimura} also calculated the progenitor mass of the Cygnus Loop and concluded that the Cygnus Loop is originated from the 12-15\MO explosion. \citet{Levenson98} also estimated the progenitor mass to be 15\MO from the size of the cavity. These results show that the progenitor star of the Cygnus Loop was a massive star and caused a core-collapse explosion. It is striking that an asymmetric explosion is suggested as the origin of the Cygnus Loop from the metal distributions. 

The Cygnus Loop is a typical shell-like SNR; this structure is thought to be generated by a cavity explosion \citep{Levenson97}. The morphology is almost circular while we can see the breakout in the south called the ``blow-out'' region \citep{Aschenbach99}. The origin of the ``blow-out'' is not well understood. \citet{Aschenbach99} have explained this extended structure as a breakout into a lower density ISM. On the other hand, \citet{Uyaniker02} suggested the existence of a secondary SNR (named G72.9-9.0) in the south based on radio observation and some other radio observations support this conclusion (\cite{Uyaniker04}; \cite{Sun06}). Recently, \citet{Uchida08} observed the blow-out region with the \textit{XMM-Newton} observatory and showed that the X-ray spectra of this region consist of a two-$kT_e$ plasma component. Judging from the plasma temperatures and the metal distributions, \citet{Uchida08} concluded that the X-ray emission is consistent with a Cygnus Loop origin and that the high- and low-$kT_{e}$ components derived from the ejecta and the cavity-wall, respectively. They also showed that the X-ray shell is thin in the blow-out region and concluded that the origin of the blow-out can be explained as a breakout into a lower density ISM as proposed by \citet{Aschenbach99}.

In this paper, we used 14 and 7 pointings observation data obtained by the \textit{Suzaku} and the \textit{XMM-Newton} observatories. We reanalyzed all the data to study the metal distributions inside the Cygnus Loop.

\section{Observations}
We summarized the 21 observations in table \ref{tab:info}. All of the Suzaku data were analyzed with version 6.5 of the HEAsoft tools. For the reduction of the \textit{Suzaku} data, we used the version 9 of the Suzaku Software. The calibration database (CALDB) used was the one updated in July 2008. We used the revision 2.2 of the cleaned event data and combined the 3$\times$3 and 5$\times$5 event files. The P03, P04, P05, P06, P07, P09, P11, and P20 data were taken by using the spaced row charge injection (SCI) method \citep{Prigozhin08} which reduces the effect of radiation damage of the XIS and recovers the energy resolution. In order to exclude the background flare events, we obtained the good time intervals (GTIs) by including only times at which the count rates are within $\pm2\sigma$ of the mean count rates.

Since the Cygnus Loop is a large diffuse source and our FOV are filled with the SNR's emission, we cannot obtain the background spectra from our FOV. We also have no background data from the neighborhood of the Cygnus Loop. We therefore applied the Lockman Hole data for the background subtraction. We reviewed the effect of the galactic ridge X-ray emission (GRXE). The flux of the GRXE at $l = 62^\circ$, $|b| < 0^\circ.4$ is $6\times10^{-12}$\,erg/cm$^2$/s/deg$^2$ (0.7-2.0\,keV) \citep{Sugizaki01}. Although the Cygnus Loop ($l = 74^\circ$, $b = -8^\circ.6$) is located outside of the FOV of \citet{Sugizaki01}, this value gives us an upper limit of the GRXE at the Cygnus Loop. Meanwhile, the total count rate of the Cygnus Loop is 1659 counts/s (0.8-2.01\,keV) \citep{Aschenbach99}, the flux is estimated to be $1.68\times10^{-10}$\,erg/cm$^2$/s/deg$^2$, assuming that the effective area of the \textit{ROSAT} HRI is 80\,cm$^2$ and that the Cygnus Loop is a circle $3^\circ.0$ in diameter. This value is consistent with the results from our FOV. Therefore, we concluded that the effect of the GRXE on the Cygnus Loop is vanishingly small. The solar wind charge exchange (SWCX) is also considered to be one of the correlates of the soft X-ray background below 1\,keV  \citep{Fujimoto07}. However, in terms of the Cygnus Loop, we consider that the SWCX is negligible because of the prominent surface brightness of the Loop. Thus, the Lockman Hole data obtained in 2006 and 2007 were applied for the background subtraction. We selected data whose observation dates were close to those of the Cygnus Loop observations. Since there were no photons above 3.0\,keV after the background subtraction, the energy ranges of 0.3-3.0\,keV and 0.4-3.0\,keV were used for XIS1 (back-illuminated CCD; BI CCD) and XIS0,2,3 (front-illuminated CCD; FI CCD), respectively \citep{Koyama07}. 

All of the \textit{XMM-Newton} data were processed with version 7.1.0 of the \textit{XMM} Science Analysis System (SAS). The current calibration files (CCFs) used were the one updated on June 2008. We used the data obtained with the EPIC MOS and pn cameras. These data were taken by using the medium filters and the prime full-window mode. We selected X-ray events corresponding to patterns 0-12 and flag = 0 for MOS 1 and 2, patterns 0-4 and flag = 0 for pn, respectively. In order to exclude the background flare events, we determined the GTIs in the same way as the \textit{Suzaku} data. After filtering the data, they were vignetting-corrected using the SAS task \textbf{evigweight}. For the background subtraction, we employed a blank-sky observations prepared by \citet{Read03} for similar reason with the \textit{Suzaku}'s case. After the background subtraction, the energy ranges of 0.3-3.0\,keV were used for each instrument.

\begin{table}
 \begin{center}
 \caption{Summary of the 21 observations}\label{tab:info}
  \begin{tabular}{lcccc}
\hline\hline
Obs. ID & Obs. Date& Coordinate (RA, DEC) & Position Angle & Effective Exposure\\
\hline
\hline
\textit{Suzaku Observations} \\
\hline
501014010  (P3) & 2007-11-14 &  20$^h$52$^m$09$^s$.9, 31$^\circ$36$^\prime$43$^{\prime\prime}$.4 & 58$^\circ$.4 & 7.5\,ksec\\
\hline
501015010  (P4) & 2007-11-14 &  20$^h$51$^m$11$^s$.8, 31$^\circ$22$^\prime$08$^{\prime\prime}$.4 & 58$^\circ$.6 & 18.3\,ksec\\
\hline
501016010  (P5) & 2007-11-15 &  20$^h$50$^m$11$^s$.3, 31$^\circ$10$^\prime$48$^{\prime\prime}$.0 & 58$^\circ$.8 & 19.3\,ksec\\
\hline
501017010  (P6) & 2007-11-11 &  20$^h$49$^m$11$^s$.3, 30$^\circ$59$^\prime$27$^{\prime\prime}$.6 & 59$^\circ$.0 & 28.7\,ksec\\
\hline
501018010  (P7) & 2007-11-12 &  20$^h$48$^m$18$^s$.7, 30$^\circ$46$^\prime$33$^{\prime\prime}$.6 & 59$^\circ$.2 & 21.0\,ksec\\
\hline
501019010  (P9) & 2007-11-12 &  20$^h$47$^m$14$^s$.2, 30$^\circ$36$^\prime$10$^{\prime\prime}$.8 & 59$^\circ$.4 & 16.2\,ksec\\
\hline
503055010  (P11) & 2008-05-09 &  20$^h$49$^m$48$^s$.7, 31$^\circ$30$^\prime$18$^{\prime\prime}$.0 & 58$^\circ$.5 & 22.2\,ksec\\
\hline
501029010  (P12) & 2006-05-09 &  20$^h$55$^m$00$^s$.0, 31$^\circ$15$^\prime$46$^{\prime\prime}$.8 & 58$^\circ$.7 & 13.2\,ksec\\
\hline
501030010  (P13) & 2006-05-10 &  20$^h$53$^m$59$^s$.3, 31$^\circ$03$^\prime$39$^{\prime\prime}$.6 & 58$^\circ$.9 & 13.9\,ksec\\
\hline
501031010  (P14) & 2006-05-12 &  20$^h$52$^m$58$^s$.8, 30$^\circ$51$^\prime$32$^{\prime\prime}$.4 & 59$^\circ$.1 & 18.2\,ksec\\
\hline
501032010  (P15) & 2006-05-25 &  20$^h$51$^m$58$^s$.6, 30$^\circ$39$^\prime$10$^{\prime\prime}$.8 & 59$^\circ$.3 & 17.4\,ksec\\
\hline
501033010  (P16) & 2006-05-22 &  20$^h$50$^m$58$^s$.8, 30$^\circ$27$^\prime$00$^{\prime\prime}$.0 & 59$^\circ$.5 & 20.0\,ksec\\
\hline
501034010  (P17) & 2006-05-22 &  20$^h$48$^m$49$^s$.7, 30$^\circ$00$^\prime$21$^{\prime\prime}$.6 & 60$^\circ$.0 & 13.9\,ksec\\
\hline
503056010  (P20) & 2008-05-10 &  20$^h$48$^m$00$^s$.0, 31$^\circ$10$^\prime$30$^{\prime\prime}$.0 & 58$^\circ$.8 & 22.5\,ksec\\
\hline
\hline
\textit{XMM-Newton Observations} \\
\hline
0082540201 (Pos-2) & 2002-12-03 & 20$^h$54$^m$07$^s$.2, 31$^\circ$30$^\prime$51$^{\prime\prime}$.1 & 241$^\circ$.7 & 14.4\,ksec\\
\hline
0082540301 (Pos-3) & 2002-12-05 & 20$^h$52$^m$51$^s$.1, 31$^\circ$15$^\prime$25$^{\prime\prime}$.7 & 241$^\circ$.7 & 11.6\,ksec\\
\hline
0082540401 (Pos-4) & 2002-12-07 & 20$^h$51$^m$34$^s$.7, 31$^\circ$00$^\prime$00$^{\prime\prime}$.0 & 241$^\circ$.7 & 4.9\,ksec\\
\hline
0082540501 (Pos-5) & 2002-12-09 & 20$^h$50$^m$18$^s$.4, 30$^\circ$44$^\prime$34$^{\prime\prime}$.3 & 231$^\circ$.4 & 12.6\,ksec\\
\hline
0082540601 (Pos-6) & 2002-12-11 & 20$^h$49$^m$02$^s$.0, 30$^\circ$29$^\prime$08$^{\prime\prime}$.6 & 241$^\circ$.7 & 11.5\,ksec\\
\hline
0405490101 (Pos-8) & 2006-05-13 & 20$^h$50$^m$32$^s$.2, 30$^\circ$11$^\prime$00$^{\prime\prime}$.0 & 69$^\circ$.9 & 6.5 \,ksec\\
\hline
0405490201 (Pos-9) & 2006-05-13 & 20$^h$49$^m$54$^s$.2, 29$^\circ$42$^\prime$25$^{\prime\prime}$.0 & 69$^\circ$.8 & 3.6 \,ksec\\
\hline
  \end{tabular}
 \end{center}
\end{table}

\section{Spectral Analysis}
\begin{figure}
  \begin{center}
    \FigureFile(100mm,0mm){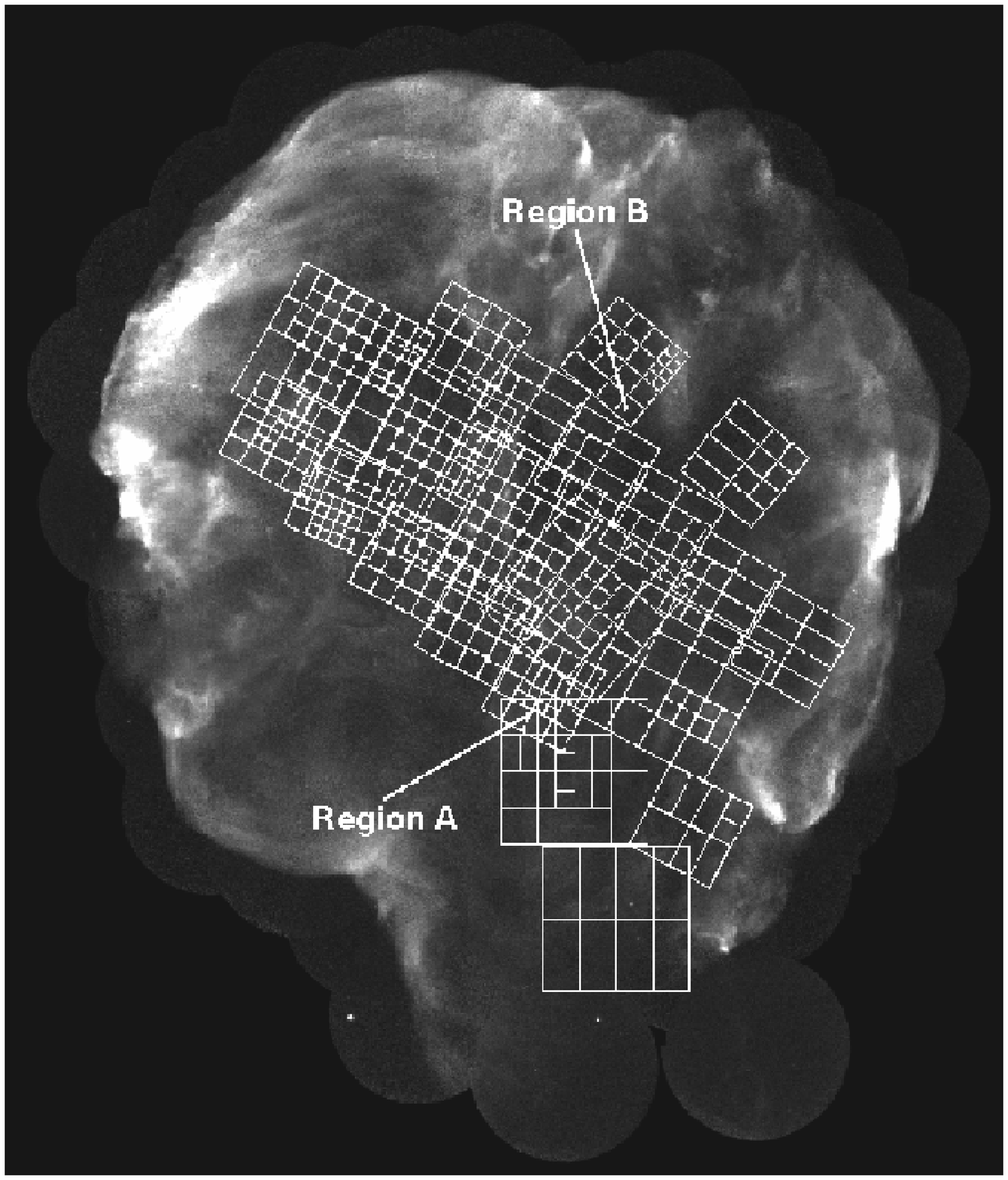}
    \FigureFile(50.5mm,0mm){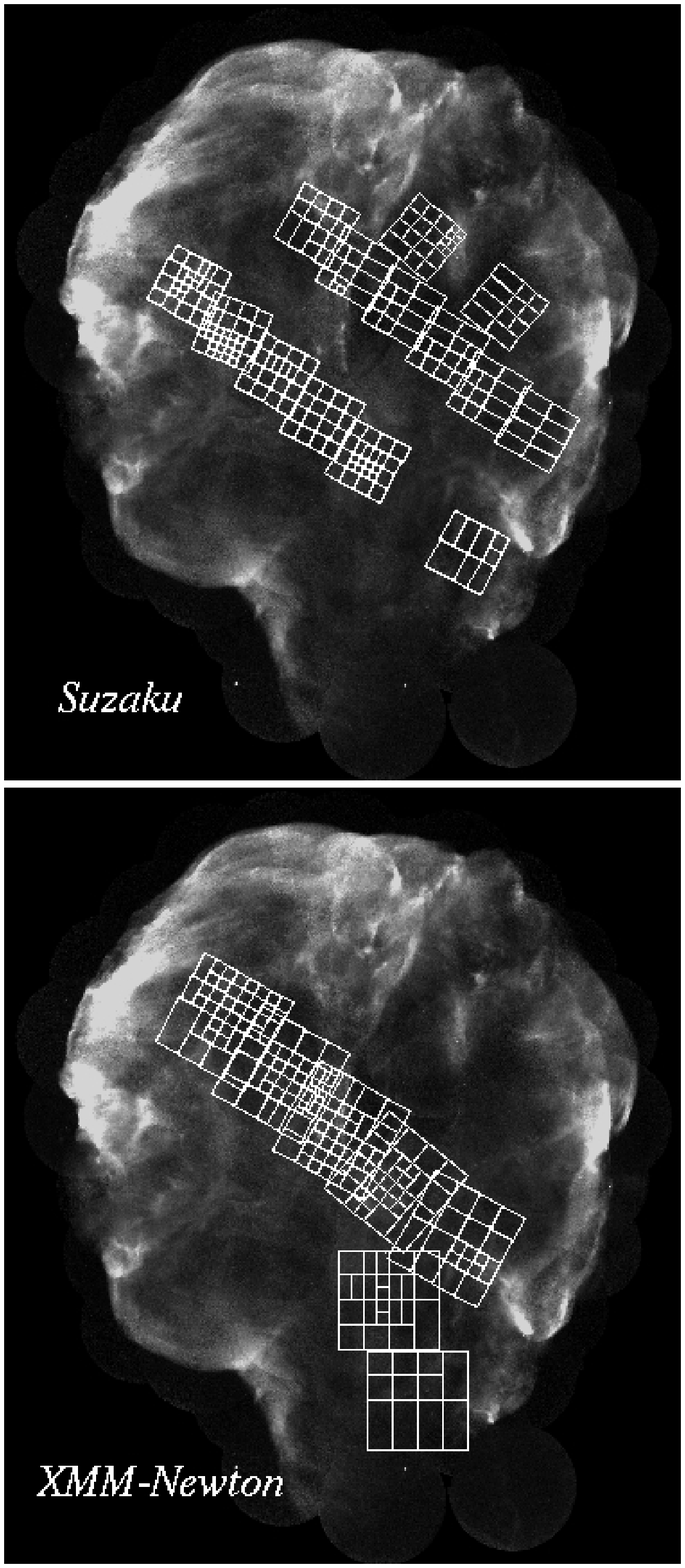}
  \end{center}
  \caption{Left panel shows the \textit{ROSAT} HRI image of the entire Cygnus Loop overlaid with the spectral extraction regions with white rectangles. Right two panels show the extraction regions for \textit{Suzaku} (top panel) and \textit{XMM-Newton} (bottom panel).}\label{fig:Region}
\end{figure}

To investigate the plasma structure of the Cygnus Loop, we divided the entire FOV into several box regions outlined in white line in Fig. \ref{fig:Region}. In order to equalize the statistics, we initially divided all images of XIS1 or MOS2 into two parts and if each divided region has more than 10,000 photons, it was once again divided. In this way, we obtained 415 box regions. Each region contains 5,000-10,000 photons for XIS1 and MOS2. The side length of each box ranges from 2.2 to 14\,arcmin. Therefore box sizes are not smaller than the angular resolution capability of the \textit{Suzaku} XIS. We grouped 415 spectra into bins with a minimum of 20 counts so that $\chi^2$ statistics are appropriate. From the earlier observation of the NE to the SW regions along the diameter, \citet{Tsunemi07} showed the plasma structure of the Cygnus Loop as follows: the high-$kT_e$ ejecta component is surrounded by the low-$kT_e$ ISM component. They found that the spectra from the inner regions of the Cygnus Loop consist of the two-component non-equilibrium ionization plasma. In order to examine the metal distributions of the ejecta of the Cygnus Loop, we need to separate the X-ray spectra into the ejecta and the surrounding cavity-material component. Then, we fitted 415 spectra extracted from our FOV by the two-$kT_e$ non-equilibrium ionization (NEI) plasma model. We employed \textbf{wabs} \citep{Morrison83} and \textbf{vnei} (NEI ver.2.0; \cite{Borkowski01}) in XSPEC version 12.4.0 \citep{Arnaud96}. In order to generate a response matrix file (RMF) and an ancillary response file (ARF), we employed xisrmfgen \citep{Ishisaki07} and xissimarfgen for the \textit{Suzaku} data, rmfgen and arfgen for the XMM-Newton data. We generated ARFs of the \textit{Suzaku} data for ``extended sources''.

First, we set all the abundances of the heavy elements free, and found that this model cannot reach the physically meaningful results. Therefore, in the low-$kT_e$ component, we fixed the metal abundances to the values from the result of \citet{Tsunemi07}. The abundances of the low-$kT_e$ ISM component shown by them are as follows: C=0.27, N=0.10, O=0.11, Ne=0.21, Mg=0.17, Si=0.34, S=0.17, Fe(=Ni)=0.20. In addition, they fixed the other elements to the solar values. Then, we used these values as the abundances of the emission from the ISM origin. As for the high-$kT_e$ component, the abundances of O, Ne, Mg, Si, and Fe were free while we set the abundances of C and N equal to O, S equal to Si, Ni equal to Fe, and other elements fixed to their solar values \citep{Anders89}. Other parameters were all free such as the electron temperature $kT_e$, the ionization timescale $\tau$ (a product of the electron density and the elapsed time after the shock heating), and the emission measure (EM $= \int n_e n_{\rm H} dl$, where $n_e$ and $n_{\rm H}$ are the number densities of hydrogen and electrons and $dl$ is the plasma depth). We also set the column density $\rm\textit{N}_H$ free.

\begin{figure}
  \begin{center}
    \FigureFile(82mm,0mm){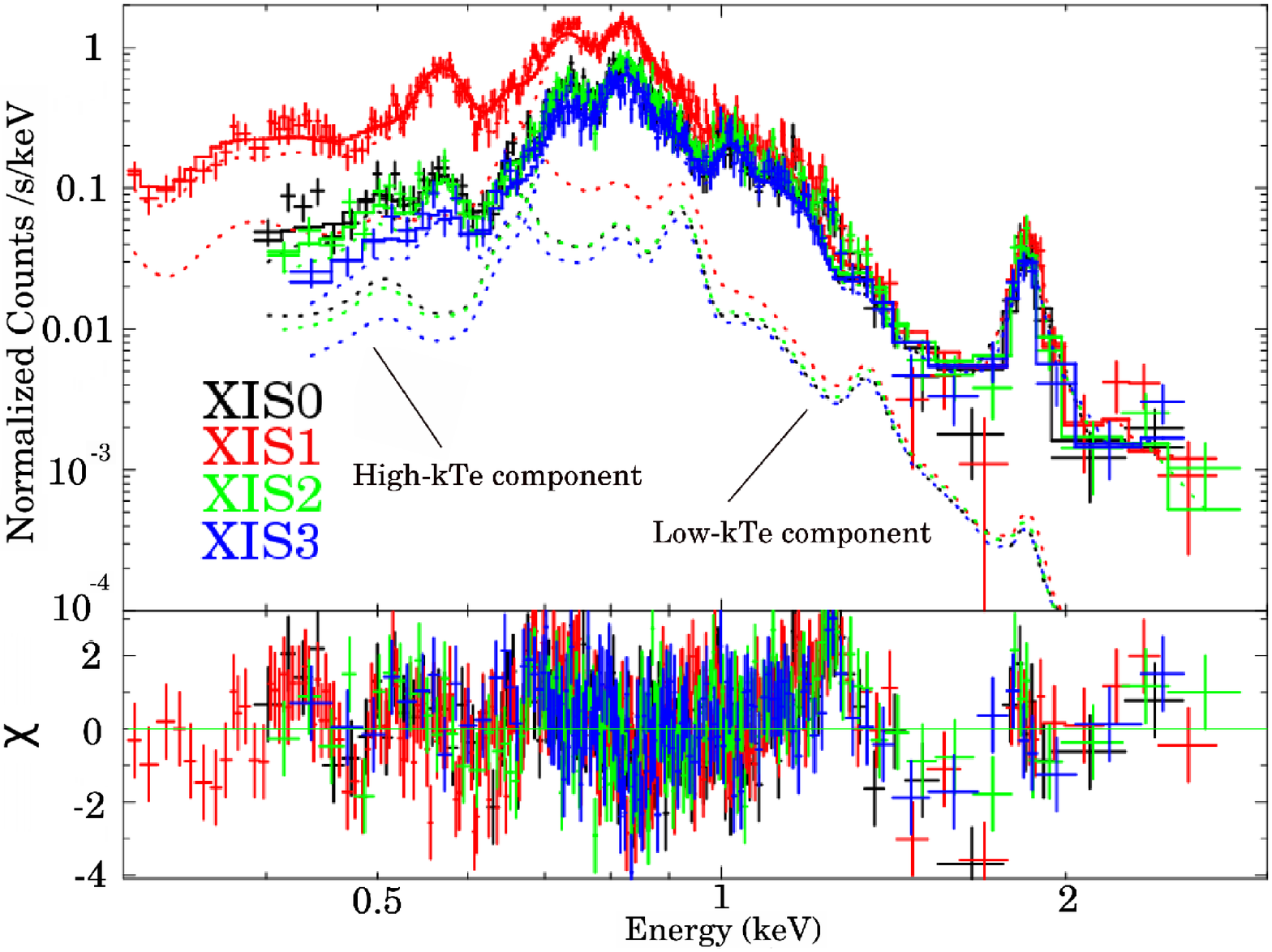}
    \FigureFile(82mm,0mm){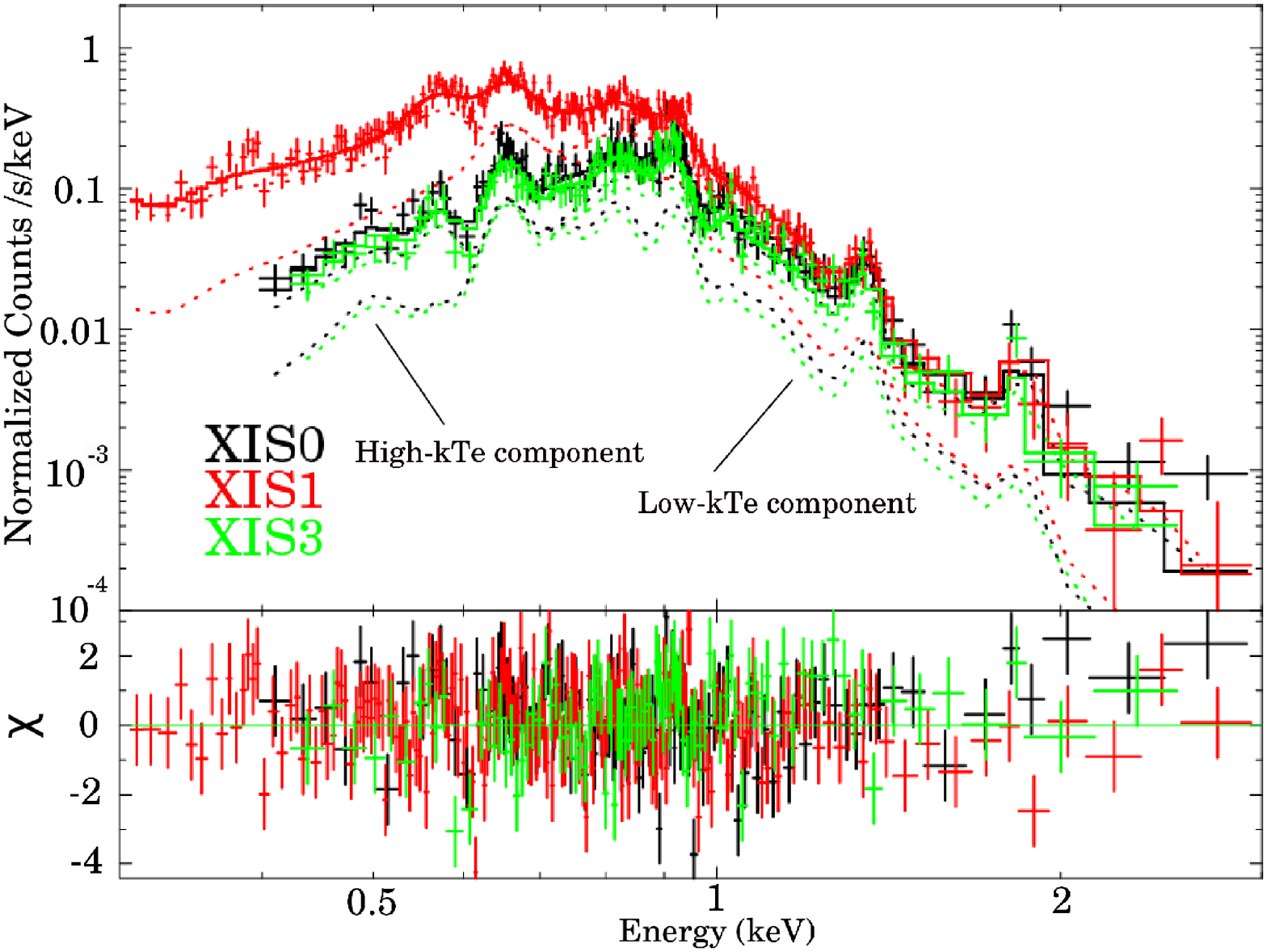}
  \end{center}
  \caption{Example spectra from the regions where Fe and Si are the most abundant (region A: left) and the most depleted (region B: right), respectively. The best-fit curves are shown with solid line and two components are shown with dotted lines. The high- and the low-$kT_e$ components cross at $\sim$0.65\,keV (region A) and $\sim$0.7\,keV (region B), respectively. The residuals are shown in the lower panels. In the left panel, black, red, green, blue correspond to  the XIS 0, 1, 2, 3. In the right panel, black, red, green correspond to the XIS 0, 1, 3.}\label{fig:spec}
\end{figure}

The spectra are reasonably well fitted by the two-$kT_e$ VNEI model for almost all regions: the values of the reduced $\chi^2$ shows around $\sim$1.5 and the degrees of freedom are 300-400. Some regions overlap each other and we fitted these spectra separately. In cases like this, we compared each best-fit parameter values and confirmed that they are just within the uncertainties. The example spectra and the best-fit curves are shown in figure \ref{fig:spec} and the best-fit parameters are shown in table \ref{tab:spec}. These two spectra are taken from the region where Fe and Si are the most abundant or the most depleted, respectively. The extracted regions are shown in figure \ref{fig:Region}. 

\begin{figure}
  \begin{center}
    \FigureFile(150mm,0mm){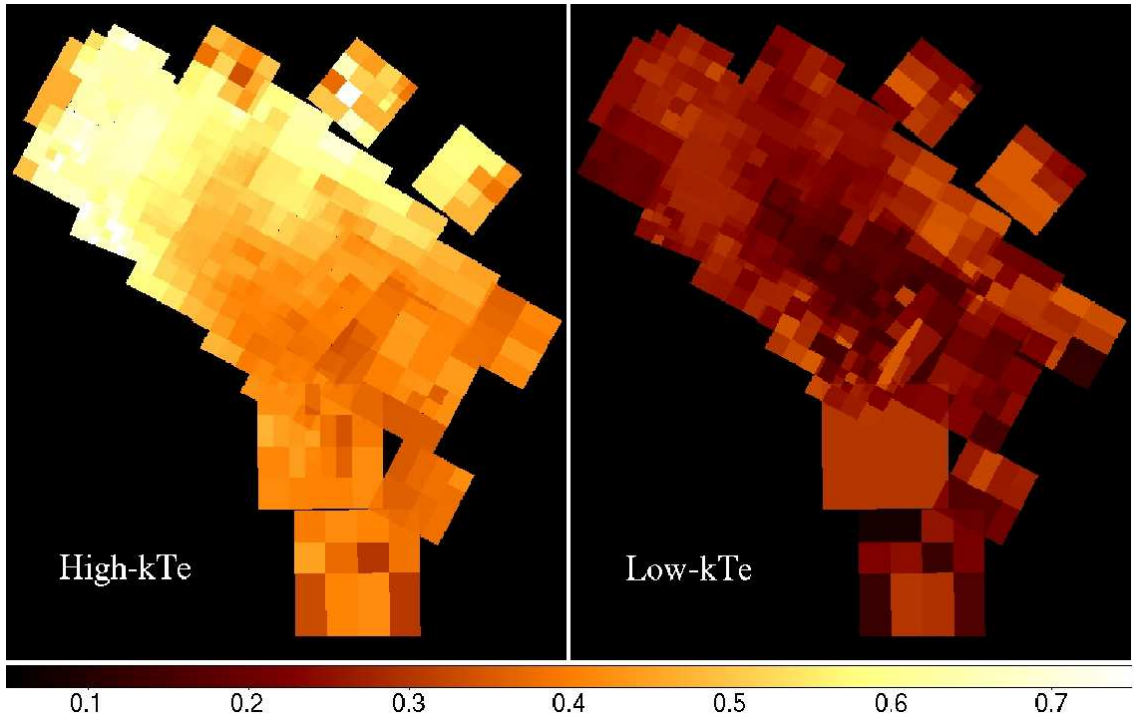}
  \end{center}
  \caption{Electron temperature distribution of each component. The left and right panel shows the distribution of the high- and low-$kT_e$ component, respectively. The values of $kT_e$ are in units of keV.}\label{fig:kTe}
\end{figure}

Figure \ref{fig:kTe} shows the electron temperature distribution of each component. The left panel shows the best-fit $kT_e$ parameters of the high-$kT_e$ component that represents the plasma temperature of the ejecta component. From the left panel, we show the inhomogeneous temperature distribution of the ejecta component. The values of $kT_e$ range from 0.4\,keV to 0.7\,keV and they are lower in the SW part than those in the NE part. The ejecta temperature of the Cygnus Loop shows smooth decrease from NE to SW. The averaged temperature is $\sim$0.49\,keV. The right panel shows the map of the $kT_e$ parameter obtained from the low-$kT_e$ component that represents the temperature of the swept-up cavity material.  The averaged value is $\sim$0.23\,keV and it ranges from 0.2\,keV to 0.3\,keV. The temperature of the low-$kT_{e}$ component near the center of the Loop is lower than that of the surrounding temperature. From figure \ref{fig:kTe}, we clearly separated the high-$kT_{e}$ component and the low-$kT_{e}$ component just as the observation obtained in \citet{Tsunemi07}, \citet{Katsuda08}, and \citet{Kimura}. 

\begin{figure}
  \begin{center}
    \FigureFile(60mm,0mm){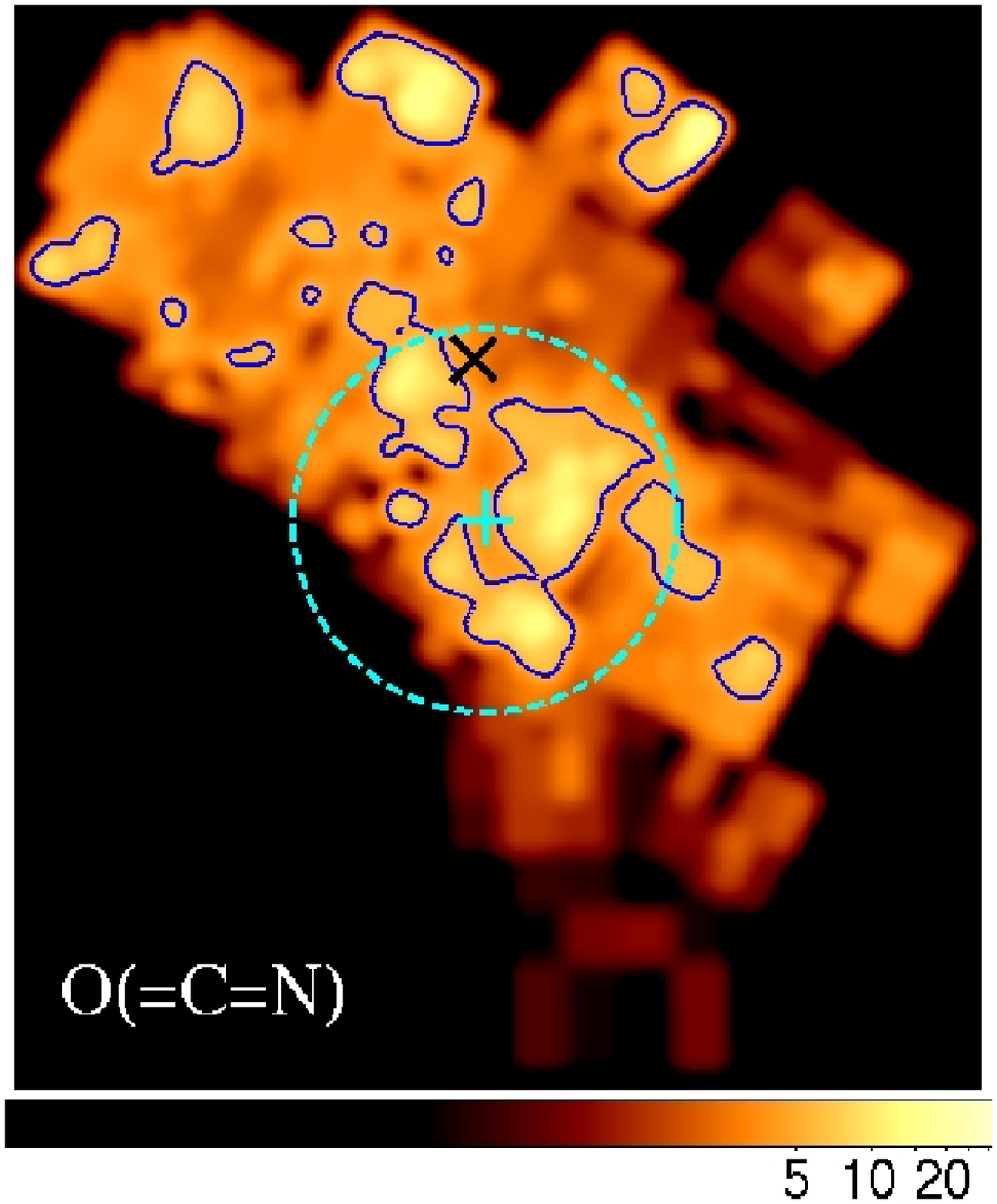}
    \FigureFile(60mm,0mm){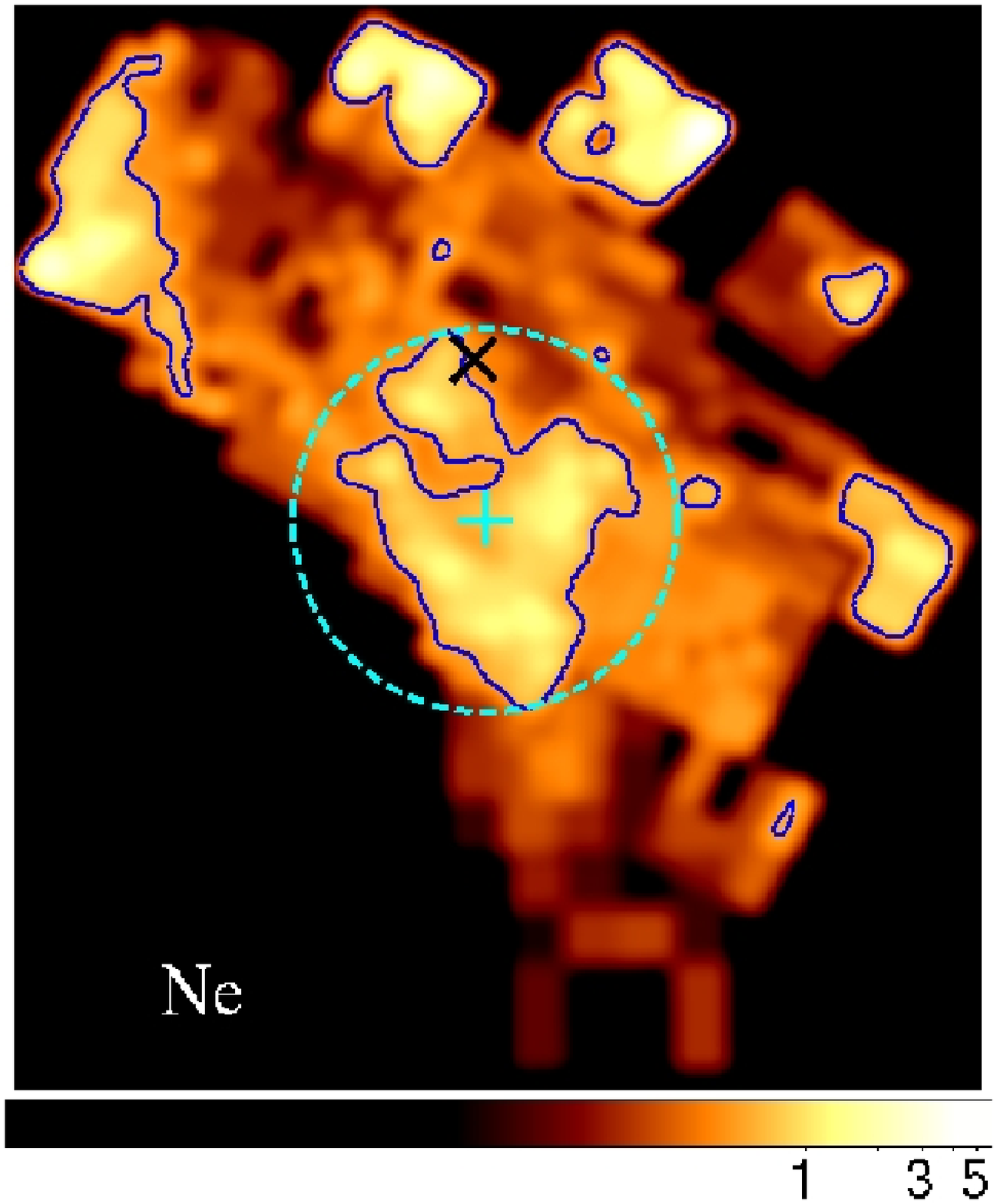}
    \FigureFile(60mm,0mm){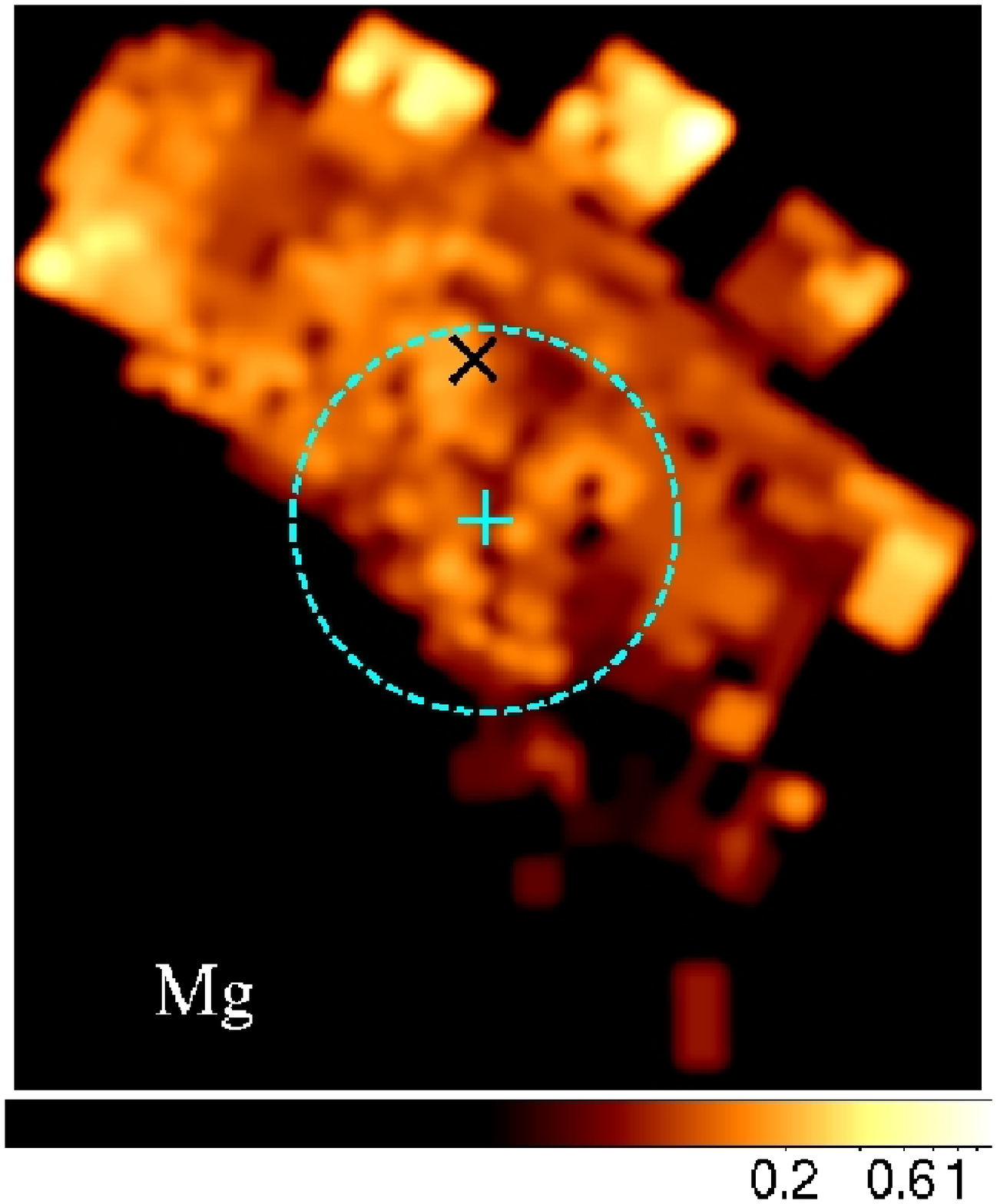}
    \FigureFile(60mm,0mm){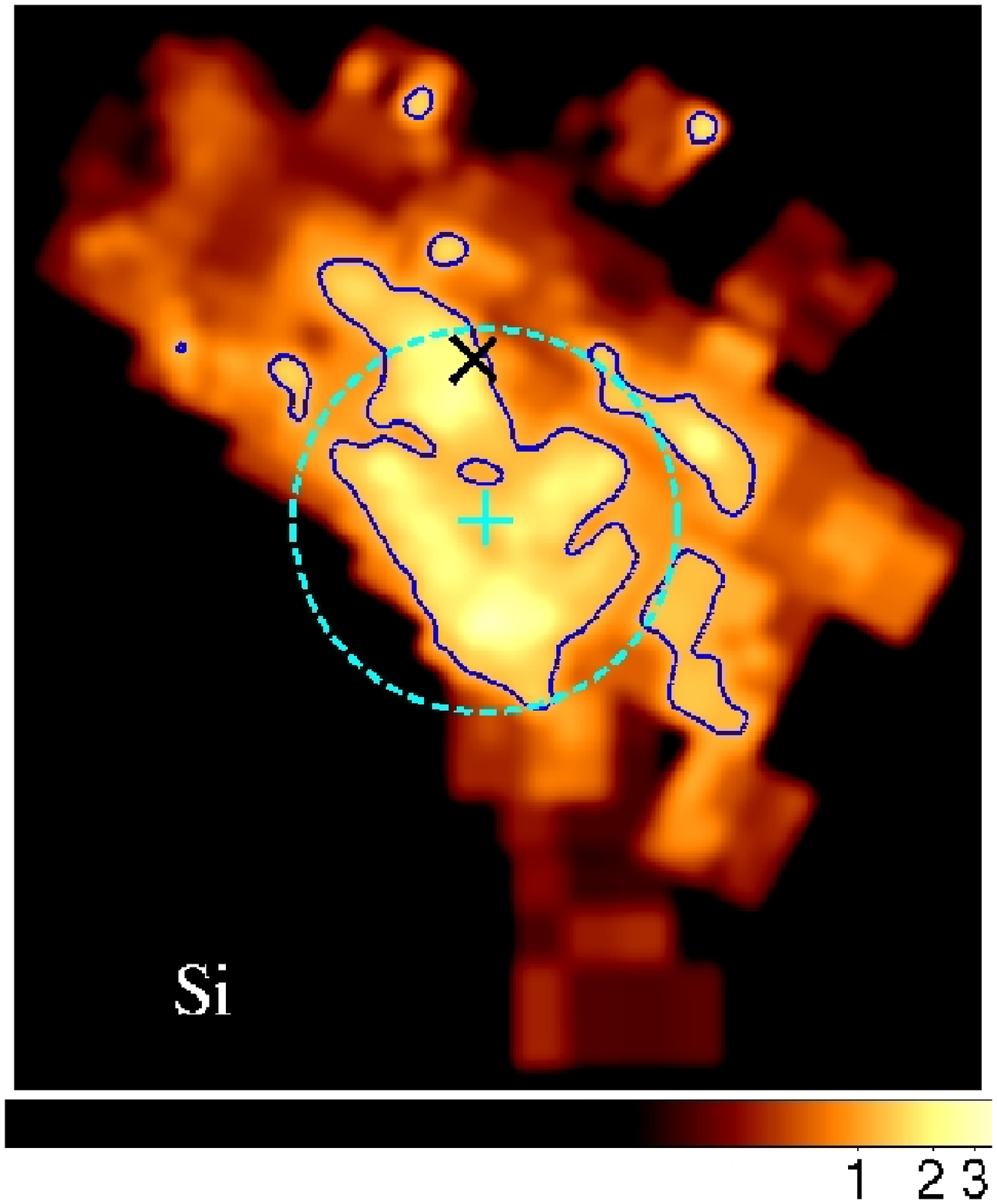}
    \FigureFile(60mm,0mm){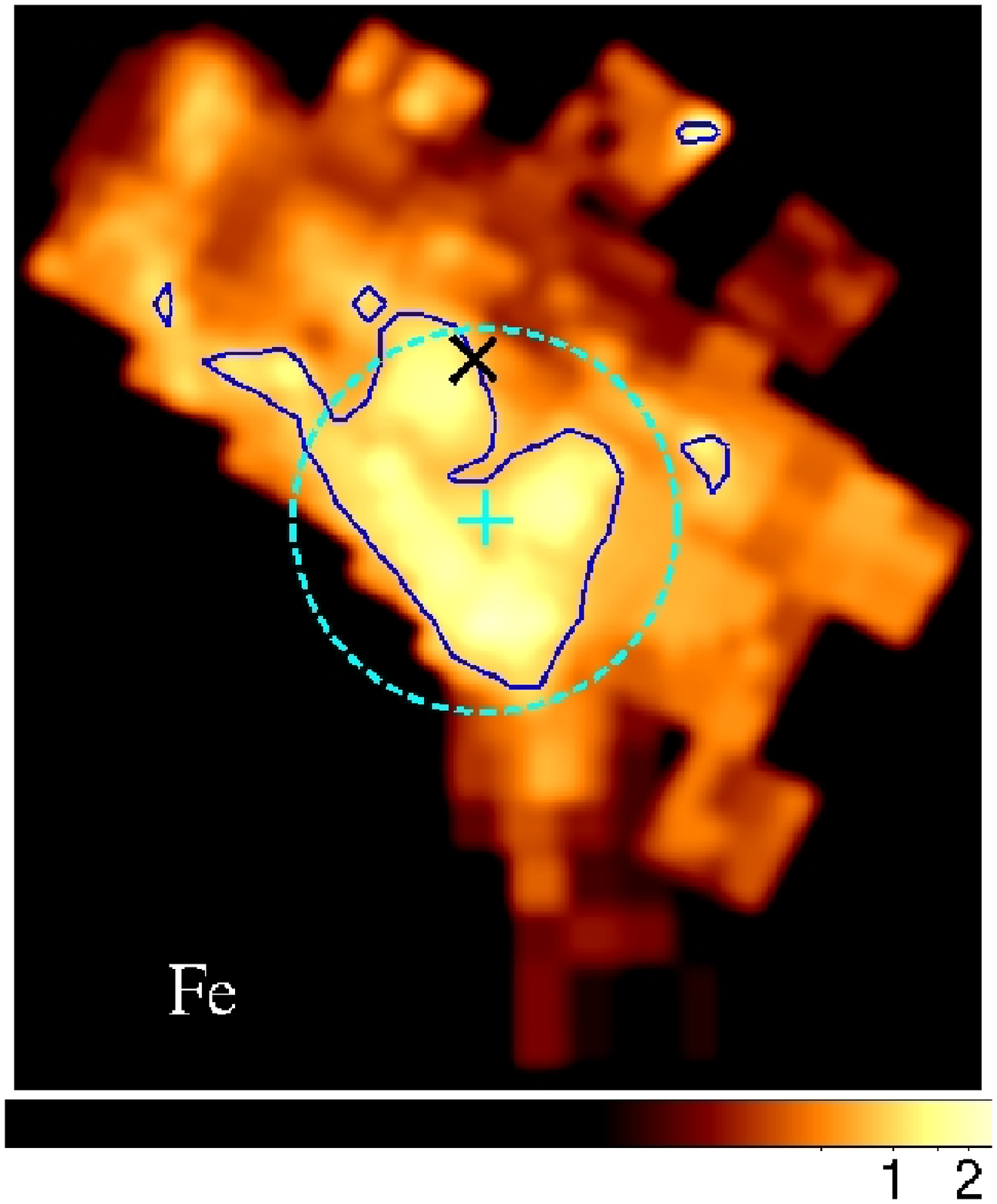}
  \end{center}
  \caption{EM distributions of the heavy elements in the logarithmic scales. The black X-mark shows the geometric center of the Loop \citep{Levenson98}. The blue contour shows the EM level of $1/e$ of the maximum around the geometric center. The light blue cross-mark and the dotted circle represent the ``metal center'' and the ``metal circle'', respectively (see \ref{sec:metal}). The values are in units of 10$^{14}$cm$^{-5}$.}\label{fig:EM}
\end{figure}

Figure \ref{fig:EM} shows the EM distributions of the heavy elements.  The images are smoothed by Gaussian kernel of $\sigma$=2.8\,arcmin.  The color code scale is normalized by the maximum values. From figure \ref{fig:EM}, the EM of Mg is relatively low compared with those of the other elements. The EM values of O, Ne, Mg are higher in the NE than those around the center while the EM values of Si and Fe are larger in the center and decrease outward. These results are consistent with those of \citet{Tsunemi07}, \citet{Katsuda08}, and \citet{Kimura}. These earlier studies were all conducted from NE to SW in their FOV. Our analysis revealed the plasma structure from the northwest (NW) to the southeast (SE) of the Loop. The averaged value of each EM [$\times$10$^{14}$\,cm$^{-5}$] is 3.20 (O=C=N), 0.39 (Ne), 0.08 (Mg), 0.81 (Si=S), 0.59 (Fe=Ni), respectively. We show the geometric center estimated by \citet{Levenson98} with the black X-mark in figure \ref{fig:EM}. It is located on $\alpha = 20^{h}51^{m}21^{s},  \delta = 31^{\circ}01^{'}37^{''}$ (J2000) which is determined by fitting the \textit{ROSAT} HRI map of the Loop by the model circle.

\begin{table}
  \begin{center}
 \caption{Spectral fit parameters}\label{tab:spec}
    \begin{tabular}{lll}
       \hline 
      \hline
 Parameter & region A & region B\\
      \hline
      N$\rm _H$ [10$^{20}$cm$^{-2}$] & 3.6 $\pm$ 0.2 & 2.6 $\pm$ 0.2 \\
      Low-$kT_e$ component: \\
      \ \ $kT_e$ [keV] & 0.19 $\pm$ 0.01 & 0.28 $\pm$ 0.01 \\ 
      \ \ C &  \multicolumn{2}{c}{0.27 (fixed)}\\
      \ \ N &  \multicolumn{2}{c}{0.10 (fixed)}\\
      \ \ O &  \multicolumn{2}{c}{0.11 (fixed)}\\
      \ \ Ne &  \multicolumn{2}{c}{0.21 (fixed)}\\
      \ \ Mg &  \multicolumn{2}{c}{0.17 (fixed)}\\
      \ \ Si &   \multicolumn{2}{c}{0.34 (fixed)}\\
      \ \ S &  \multicolumn{2}{c}{0.17 (fixed)}\\
      \ \ Fe(=Ni) &  \multicolumn{2}{c}{0.20 (fixed)}\\
      \ \ log $\tau$ & 11.49 $^{+ 0.05}_{- 0.04}$  & 11.12 $^{+ 0.03}_{- 0.03}$\\
      \ \ EM [10$^{18}$cm$^{-5}$] & 6.23 $\pm$ 0.19 & 4.27 $\pm$ 0.14\\
      High-$kT_e$ component: \\
      \ \ $kT_e$ [keV] & 0.42 $\pm$ 0.01  & 0.63 $\pm$ 0.01\\ 
      \ \ O(=C=N) & 0.47 $\pm$ 0.08 & 0.31 $\pm$ 0.02\\
      \ \ Ne & 0.75 $\pm$ 0.09  & 0.47 $\pm$ 0.05\\
      \ \ Mg & 0.21 $\pm$ 0.10 & 0.41 $\pm$ 0.08\\
      \ \ Si(=S) & 5.36 $\pm$ 0.48 & 0.49 $\pm$ 0.18\\
      \ \ Fe(=Ni) & 2.92 $\pm$ 0.05 & 0.55 $\pm$ 0.03\\
      \ \ log $\tau$ & 11.76 $\pm$ $^{+ 0.10}_{- 0.06}$ & 10.82 $\pm$ 0.02\\
      \ \ EM [10$^{18}$cm$^{-5}$] & 0.55 $\pm$ 0.01 & 0.68 $\pm$ 0.02\\
$\chi ^2$/dof & 441/308 & 527/402\\
      \hline
    \end{tabular}
 \end{center}
\end{table}

\section{Discussion}
\subsection{Low abundances of the low-$kT_e$ component}\label{sec:low}
We analyzed all the spectra with using a two-component model: the low-$kT_e$ component forms a shell-like structure and the high-$kT_e$ component fills inside (\cite{Tsunemi07}; \cite{Katsuda08}; \cite{Kimura}). The fixed abundances of the low-$kTe$ component are depleted compared to the solar values. \citet{Cartledge04} measured the abundance of the interstellar oxygen in the solar neighborhood along 36 sight lines and showed 0.4 times the solar value. \citet{Wilms00} employed 0.6 of the total interstellar abundances for the ISM oxygen abundance. The abundances we fixed are significantly lower than those results and it needs to be discussed. The metal deficiency of the ISM component of the Cygnus Loop has been shown by the earlier X-ray observations \citep{Miyata99}. The other observations in X-ray by using the \textit{Suzaku}, \textit{XMM-Newton}, and \textit{Chandra} also showed the metal depletion at the rim (\cite{Leahy04}; \cite{Tsunemi07}; \cite{Katsuda08b}). Meanwhile, \citet{Katsuda08b} found an enhanced abundance region in the very edge of the NE rim.  The abundance there is about half the solar value which is much higher than those other rim regions.  This result is consistent with the observations of the ISM abundance in the solar neighborhood (\cite{Wilms00}; \cite{Cartledge04}). Therefore, the metal abundances of the periphery of the Loop may be higher than those observed in X-ray. Any line of sight through the remnant almost certainly intersects various emitting regions. However, the line of sight near the limb can intersect the very limited plasma condition. Therefore, we can expect that the very rim region \citet{Katsuda08b} observed consists of thermal plasma with ISM abundance while other regions may include either thin thermal plasma with low abundance or non-thermal emission. \citet{Katsuda08c} tried to fit the spectrum with a combination of thin thermal plasma with ISM abundance and the non-thermal emission.  They reached no clear result probably due to the simplicity of the model.  Although most of the rim regions show low abundance from the X-ray data analysis point of view, it still remains an open question.

\subsection{Displacement of the metal distribution from the geometric center}\label{sec:metal}

 We noticed that the strong EM regions concentrate just south of the geometric center.  Si and Fe concentrate in the inner part while Ne and O have strong regions both in the inner part and the outer part within our FOV.  Mg has strong regions only in the outer part.  In this paper, we will focus on the strong regions near the geometric center.

The contours in figure \ref{fig:EM} show the EM level of $1/e$ of the maximum near the geometric center.  The concentration
of the inner part of various metals is obviously away from the geometric center.  The spread of the strong regions indicated by the
contour is not clear in the SE region due to the lack of observation.  Therefore, it is difficult to precisely determine the spread. Taking into account the lack of  information in the SE region, we speculate that the center of the metal distribution is determined by the centroid of the Fe distribution at the south of the geometric center. We call it the ``metal center'', which is located on $\alpha = 20^{h}51^{m}14^{s},  \delta = 30^{\circ}40^{'}35^{''}$ (J2000), as shown by the light blue  cross-mark in figure \ref{fig:EM}. 

We found that the metal center is separated from the geometric center by 25 arcmin toward the south. It is not clear why it is displaced from the geometric center. On the simple assumption that the SN explosion occurred symmetrically in the uniform ISM, the metals should distribute symmetrically and its center should overlap with the geometric center. The Cygnus Loop shows a nearly circular shape with an exception of the south blow-out region. However, the X-ray surface brightness and the optical filamentation suggest that the ambient medium of the Cygnus Loop is not uniform. We consider that its non-uniformity is prominent toward the south blow-out region. If the south blow-out is originated from the lower-density cavity wall or the ISM, the displacement of the metal distributions to the south could be naturally explained. In general, the non-uniformity produces the asymmetrical reverse shocks.  Taking into account the estimated age of the Loop, the reverse shocks should have reached the center of the Loop. However, the times for the reverse shocks to return to the center, $t_\mathrm{R}$ significantly depend on the ambient density in the direction which the forward shock propagates. From the recent theoretical calculation, \citet{Ferreira08} obtained the relation of $t_\mathrm{R} \propto (\rho_\mathrm{ISM})^{-1/3}$, where $\rho_\mathrm{ISM}$ represents the ambient density. Therefore, in directions where $\rho_\mathrm{ISM}$ is higher, the forward shock decelerates and the reverse shock propagates toward the center more quickly. From the morphological point of view, it is suggested that the cavity wall must be very thin in the south blow-out region, although the origin of the blow-out is still not clear.

As described in section \ref{sec:intro}, from the radio observations, \authorcite{Uyaniker02} (2002; 2004) and \citet{Sun06} claim that the extra SNR exists in the south and interacts with the Cygnus Loop. Their main arguments are the difference of the radio morphology and the polarization intensity between the main part of the Loop and the south blow-out. However, based on the X-ray data, \citet{Uchida08} obtained no evidence that a smaller SNR exists at the same distance to the Cygnus Loop. They showed that the X-ray spectra in the south are interpretable by the Cygnus Loop origin. They also found that the emission from the cavity material of the Loop is relatively weak. It suggests that the X-ray shell is very thin in the south. On the other other hand, the ambient medium in the north part of the Cygnus Loop is thought to be denser than that in the south. The north areas have the strong emission in X-ray and it suggests that the forward shock has decelerated enough to become radiative in places. Then, the reverse shocks in some part other than south should be formed earlier and reaches the center faster than that in the south. In that case, the pressure exerted by the north reverse shock is higher than that exerted by the south one. If the metal distributions are formed by the imbalance of the pressure surrounding them, they should be created southward from the geometric center. Our result is consistent with the assumption that the south blow-out is originated not from an extra SNR but from a non-uniformity of the cavity wall.

\subsection{Metal circle and the progenitor star mass}
\begin{figure}
  \begin{center}
    \FigureFile(120mm,0mm){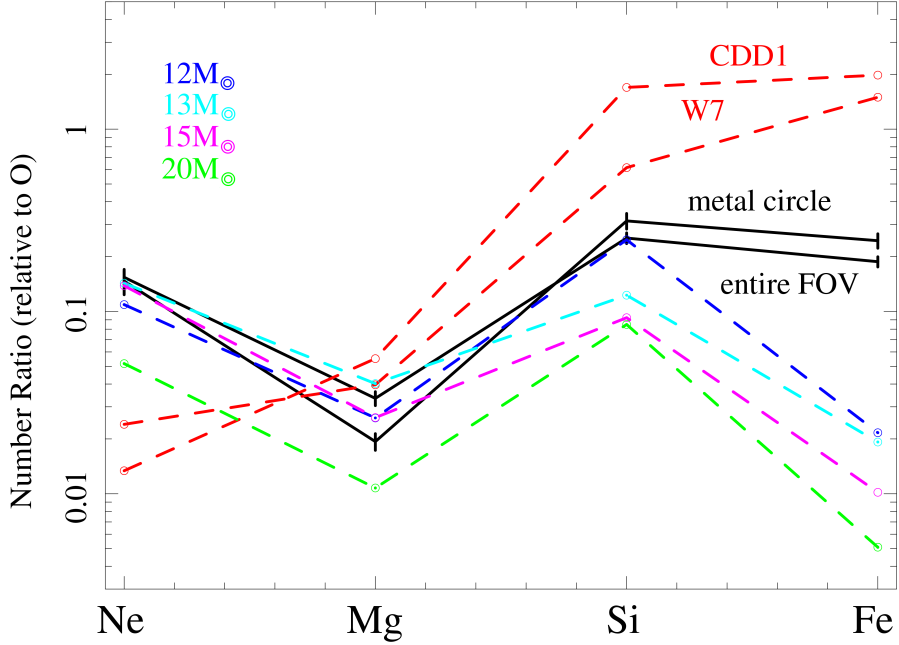}
 \end{center}
  \caption{Number ratios of Ne, Mg, Si, and Fe relative to O of the high-$kT_e$ component estimated from the metal circle and that from the entire FOV (black lines). Dotted red lines represent the CDD1 and W7 Type Ia supernova models of Iwamoto et al. (1999). Dotted blue, light blue, magenta, and green lines represent core-collapse models with progenitor masses of 12, 13, 15, 20\MO, respectively \citep{Woosley95}.}\label{fig:ratio}
\end{figure}

Although the metal abundance is not uniform within the high-$kT_e$ component, the contents in our FOV reflect the abundance of the progenitor star.  We calculate the metal abundances around the metal center and in our FOV relative to O in order to compare them with the theoretical models. We defined the ``metal circle'' which is centered in the metal center with radius of 25\,arcmin. We added the EM of ejecta component of each region weighting each area of the region for the heavy elements, O, Ne, Mg, Si, and Fe. These values reflect the total amount of each heavy element in our FOV. Then, we obtained the number ratios of each element relative to O. Figure \ref{fig:ratio} shows the number ratios of Ne, Mg, Si and Fe relative to O of the ejecta component. The vertical bars show the uncertainties of the values. The black lines represent the results from the metal circle and the entire FOV. We also plotted  the core-collapse models \citep{Woosley95} for various progenitor masses and Type Ia SN models \citep{Iwamoto99} for comparison. The result from the metal circle is similar to that from our entire FOV, which suggests the mixing of the metal. We found that  the calculated Ne/O, Mg/O, and Si/O are in good agreement with the core-collapse model with the progenitor mass of 12\MO. However, the Fe/O ratio is 10 times higher than that of the 12\MO model, and it does not fit any other models. This may be partly because our FOV does not cover the entire region of the Loop. From figure \ref{fig:EM}, we found that the EM values of O, Ne, Mg are higher around our FOV rather than those near the center of the Loop. By contrast, Si and Fe decrease toward the outside of the Loop. If these trend continue to the outside of our FOV, we expect that the total EMs of O, Ne, Mg will increase considerably more than those of Si and Fe. In that case, Si/O and Fe/O estimated by the whole remnant will decrease and approach the values the core-collapse models rather than those of the Type Ia SN models. We can conclude that the progenitor star mass of the Cygnus Loop is 12-15\MO.

\section{Conclusion}
We analyzed the metal distribution of the Cygnus Loop using 14 and 7 pointings observation data obtained by the \textit{Suzaku} and the \textit{XMM-Newton} observatories. We fit all the spectra by the two-$kT_e$ non-equilibrium ionization plasma model (VNEI model) as shown by the earlier observations (\cite{Tsunemi07}; \cite{Katsuda08}; \cite{Kimura}).  The results indicate that Si and Fe are more abundant near the center of the Loop than those around our FOV. We measured a center of the Si and Fe distributions from the metal spread and called it as ``metal center''. We found that the metal center is located at the southwest of the geometric center toward the blow-out region.

 From the best-fit parameters, we also estimated the progenitor mass of the Cygnus Loop. We calculated the number ratios of the heavy elements (relative to O) both from the entire FOV and inside the metal circle. The result from the metal circle is not different from that of our entire FOV, which suggests mixing of the metals. The results show that Ne/O, Mg/O, and Si/O are well fitted by the core-collapse models with progenitor masses of 12, 13, 15\MO. However, Fe/O is 10 times higher than those of the models. This may be partly because our FOV does not cover the entire region of the Loop. We can conclude that the progenitor mass of the Cygnus Loop is 12-15\MO.

\section*{Acknowledgements}

This work is partly supported by a Grant-in-Aid for Scientific Research by the Ministry of Education, Culture, Sports, Science and Technology (16002004).  This study is also carried out as part of the 21st Century COE Program, \lq{\it Towards a new basic science: depth and synthesis}\rq. H.U. and S.K. are supported by JSPS Research Fellowship for Young Scientists.


\begin{thebibliography}{}
\bibitem[Anders \& Grevesse (1989)]{Anders89}
Anders, E., \& Grevesse, N. \ 1989, Geochim. Cosmochim. Acta, 53, 197
\bibitem[Arnaud (1996)]{Arnaud96}
Arnaud, K. A. \ 1996, in ASP Conf. Ser. 101, Astronomical Data Analysis Software and Systems V, ed. G. H. Jacoby \& J. Barnes (San Francisco: ASP), 17
\bibitem[Aschenbach \& Leahy (1999)]{Aschenbach99}
Aschenbach, B., \& Leahy, D. A.\ 1999, \aap, 341, 602
\bibitem[Blair \etal (2005)]{Blair05}
Blair, W. P., Sankrit, R., \& Raymond, J. C. \ 2005, AJ, 129, 2268
\bibitem[Borkowski \etal (2001)]{Borkowski01}
Borkowski, K. J., Lyerly, W. J., \& Reynolds, S. P. \ 2001, \apj, 548, 820
\bibitem[Cartledge \etal (2004)]{Cartledge04}
Cartledge, S. I. B., Lauroesch, J. T., Meyer, D. M., \& Sofia, U. J. \etal \ 2004, \apj, 613, 1037
\bibitem[Ferreira \& de Jager (2008)]{Ferreira08}
Ferreira, S. E. S., \& de Jager, O. C. \ 2008, \aap, 478, 17
\bibitem[Fujimoto \etal (2007)]{Fujimoto07}
Fujimoto, R., \etal \ 2007, \pasj, 59, S133
\bibitem[Hatsukade \& Tsunemi (1990)]{Hatsukade90}
Hatsukade, I. \& Tsunemi, H. \ 1990, \apj, 362, 566
\bibitem[Ishisaki \etal (2007)]{Ishisaki07}
Ishisaki, Y., \etal \ 2007, \pasj, 59, S113 
\bibitem[Iwamoto \etal (1999)]{Iwamoto99}
Iwamoto, K., Brachwitz, F., Nomoto, K., Kishimoto, N., Umeda, H., Hix, W. R., \& Thielemann, F.-K. 1999, \apjs, 125, 439
\bibitem[Katsuda \etal (2008a)]{Katsuda08}
Katsuda, S., \etal \ 2008a, \pasj, 60, SP1, S107
\bibitem[Katsuda \etal (2008b)]{Katsuda08b}
Katsuda, S., \etal \ 2008b, \pasj, 60, SP1, S115
\bibitem[Katsuda \etal (2008c)]{Katsuda08c}
Katsuda, S., \etal \ 2008c, \apj, 680, 1198
\bibitem[Kimura \etal (2009)]{Kimura}
Kimura, M., Tsunemi, H., Katsuda, S. \& Uchida, H. \ 2008, \pasj
\bibitem[Koyama \etal (2007)]{Koyama07}
Koyama, K. \etal \ 2007, \pasj, 59S, 23
\bibitem[Leahy (2004)]{Leahy04}
Leahy, D. A. \ 2004, MNRAS, 351, 385
\bibitem[Levenson \etal (1997)]{Levenson97}
Levenson, N. A., Graham, J. R., \& Walters, J. L. \ 1997, \apj, 484, 304
\bibitem[Levenson \etal (1998)]{Levenson98}
Levenson, N. A., Graham, J. R., Keller, L. D., \& Richter, M. J. \ 1998, \apjs, 118, 541
\bibitem[Miyata \etal (1998)]{Miyata98}
Miyata, E., Tsunemi, H., Kohmura, T., Suzuki, S., \& Kumagai, S. \ 1998, \pasj, 50, 257
\bibitem[Miyata \& Tsunemi (1999)]{Miyata99}
Miyata, E., \& Tsunemi, H. \ 1999, \apj, 525, 305
\bibitem[Morrison \& McCammon (1983)]{Morrison83}
Morrison, R. \& McCammon, D. \ 1983, \apj, 270, 119
\bibitem[Prigozhin \etal (2008)]{Prigozhin08}
Prigozhin, G., Burke, B., Bautz, M., Kissel, S., LaMarr, B. \ 2008, IEEE Transactions on Electron Devices, 55, 2111
\bibitem[Read \& Ponman (2003)]{Read03}
Read, A. M., \& Ponman, T. J. \ 2003, \aap, 409, 395
\bibitem[Sugizaki \etal (2001)]{Sugizaki01}
Sugizaki, M., Mitsuda, K., Kaneda, H., Matsuzaki, K., Yamaguchi, S., \& Koyama K. \ 2001, \apj, 134, 77
\bibitem[Sun \etal (2006)]{Sun06}
Sun X. H., Reich, W., Han, J. L., Reich, P., \& Wielebinski, R. \ 2006, \aap, 447, 937
\bibitem[Tsunemi \etal (1988)]{Tsunemi88}
Tsunemi, H., Manabe, M., Yamashita, K., \& Koyama, K. \ 1988, \pasj, 40, 449
\bibitem[Tsunemi \etal (2007)]{Tsunemi07}
Tsunemi, H., Katsuda, S., Norbert, N., \& Miller, E. D. \ 2007, \apj, 671, 1717
\bibitem[Uchida \etal (2008)]{Uchida08}
Uchida, H., Tsunemi, H., Katsuda, S. \& Kimura, M. \ 2008, \apj, 688, 1102
\bibitem[Uyaniker \etal (2002)]{Uyaniker02}
Uyaniker, B., Reich, W., Yar, A., Kothes, R., \& Furst, E. \ 2002, \aap, 389, L61
\bibitem[Uyaniker \etal (2004)]{Uyaniker04}
Uyaniker, B., Reich, W., Yar, A., \& Furst, E. \ 2004, \aap, 426, 909
\bibitem[Wilms \etal (2000)]{Wilms00}
Wilms, J., Allen, A., \& McCray, R. \ 2000, \apj, 542, 914
\bibitem[Woosley \& Weaver (1995)]{Woosley95}
Woosley, S. E., \& Weaver, T. A. \ 1995, \apjs, 101, 181
\end{thebibliography}
\end{document}